\documentstyle[12pt]{article}

\begin{document}

\author{R. Vilela Mendes \\
Grupo de F\'\i sica-Matem\'atica\\
Complexo II, Universidade de Lisboa \\
Av. Gama Pinto, 2, 1699 Lisboa Codex Portugal\\
e-mail: vilela@alf4.cii.fc.ul.pt}
\title{Saddle scars: Existence and applications}
\date{}
\maketitle

\begin{abstract}
A quantum scar is a wave function which displays an high intensity in the
region of a classical unstable periodic orbit. Saddle scar are states
related to the unstable harmonic motions along the stable manifold of  a
saddle point of the potential. Using a semiclassical method it is shown
that, independently of the overall structure of the potential, the local
dynamics of the saddle point is sufficient to insure the general existence
of this type of scars and their factorized structure is obtained.
Potentially useful situations are identified, where these states appear
(directly or in disguise) and might be used for quantum control purposes.
\end{abstract}

\section{Introduction}

Until the early eighties it was widely believed that, for systems with an
ergodic classical motion, the squared eigenfunctions must coincide, in the
semiclassical limit, with the projection of the microcanonical phase space
measure. This idea found solid ground on the mathematical results of
Shnirelman\cite{Shnirelman}, Zelditch\cite{Zelditch} and Colin de Verdi\`ere%
\cite{Colin}. On close scrutiny however, what these results state is that,
for a quantum system that is classically ergodic, there is an eigenvalue
sequence of density one such that the corresponding quantum densities $%
\left| \psi (x)\right| ^2$ converge weakly to the Liouville measure.
Therefore the observation of states that do not fit these expectations does
not contradict the mathematical results. For one thing the convergence may
be very slow and, on the other hand, nothing forbids the existence of other
subsequences converging to measures different from the Liouville measure.

In fact wave functions were found which are concentrated near the classical
unstable periodic orbits. When this happens one says that {\it the quantum
state is scarred by the unstable periodic orbit} or that one has a {\it %
quantum scar}. Such states have been observed at first in numerical
simulations\cite{Taylor} \cite{Heller1} \cite{Heller2} and, more recently,
experimental evidence was found\cite{Wilkinson} on a semiconductor
quantum-well tunneling experiment.

The first theory of scars was proposed by Heller\cite{Heller1}, other
theoretical formulations followed, developed by several authors\cite
{Bogomolny} \cite{Berry} \cite{Feingold1} \cite{Feingold2}. Heller's theory
studies the overlap integral 
\begin{equation}
\label{1.1}C(t)=\left\langle \Psi (t,x)\mid \Psi (0,x)\right\rangle 
\end{equation}
for a propagating wave packet which at time zero has a Gaussian shape and
initial conditions $(p_0,x_0)$ corresponding to an unstable periodic orbit.
Expanding $\Psi (0,x)$ in energy eigenstates 
\begin{equation}
\label{1.2}\Psi (0,x)=\sum_nc_n\Psi _n(x) 
\end{equation}
one sees that the Fourier transform $S(E)$ of the overlap $C(t)$ is the
spectral density weighted by the probabilities $\left| c_n\right| ^2$. 
\begin{equation}
\label{1.3}S(E)=\sum_n\left| c_n\right| ^2\delta (E-E_n) 
\end{equation}
Now, if the period $\tau $ of the classical periodic orbit and the largest
positive Lyapunov exponent $\lambda $ are such that $e^{-\tau \lambda /2}$
is not very small, the overlap $C(t)$ will display peaks at times $n\tau $.
As the wave packet spreads, the amplitude of the peaks decreases after each
orbit traversal at the rate $e^{-\tau \lambda /2}$. The Fourier transform of 
$C(t)$ will therefore have peaks of width $\lambda $ with spacing $\omega =%
\frac{2\pi }\tau $. Referring to (\ref{1.3}) one concludes that only the
eigenstates that lie under the peaks contribute to the expansion of the wave
packet. Since the wave packet has an enhanced intensity along the region of
the period orbit, this is expected to carry over to the contributing energy
eigenstates. The stronger the overlap resurgences are, the stronger the
effect is expected to be. Therefore the intensity of the effect varies like $%
1/\tau \lambda $ .

The above qualitative derivation\cite{Heller1} of the scar effect is flawed
if the product $\lambda d(E)$ (where $d(E)$ is the mean level density) is
very large. Then the number of contributing eigenstates is very large and no
individual eigenstate is required to show a significant intensity
enhancement near the periodic orbit. Also the argument assumes the low
period unstable orbits to be isolated. If there are several nearby orbits of
different periods the argument breaks down. However, if it happens that many
periodic orbits of the same period are present in the same configuration
space region, the effect may even be enhanced. This is the situation for the
periodic motions in the neighborhood of an unstable critical point (a
saddle) of the potential $V(x)\,$. Near the critical point there are
unstable harmonic periodic motions along the stable manifold of the critical
point. As long as anharmonic corrections are unimportant, all the orbits
will have the same period independently of their amplitude. The scars
associated to these unstable periodic orbits are called {\it saddle scars}
in this paper.

Whenever a dynamical systems has a phase-space region with sensitive
dependence to initial conditions, the periodic orbits in that region are
unstable and, even when they are dense, they are a zero measure set in the
smooth (Liouville) measure over the energy surface. Therefore, unstable
classical orbits are in practice never observed, because all typical motions
are aperiodic and uniformly cover the support of the Liouville measure. The
phenomenon of quantum scars may therefore have far-reaching implications for
the applications of quantum systems. Whenever an unstable periodic orbit
scars a quantum eigenstate, the system may easily be made to behave like the
unstable orbit by resonant excitation to the corresponding energy level. In
this sense, scars are a gift of Nature, for they allow the exploration of
dynamical configurations that in classical mechanics are washed away by
ergodicity.

Saddle points are the typical critical points of generic (Morse) functions.
Therefore, once their existence is established, saddle scars are expected to
be quite abundant. In the remainder of the paper a semiclassical method is
used to establish that, independently of the overall structure of the
potential, the local dynamics of the saddle point is sufficient to insure
the existence of this type of scars. Then, in the closing section,
potentially useful situations are identified where these states appear
(directly or in disguise) and may be used for quantum control purposes.

\section{Semiclassical estimates}

In the neighborhood of a saddle point, there is a choice of coordinates such
that, up to higher order terms, the potential is 
\begin{equation}
\label{2.1}V(x)=\sum_i\sigma _ix_i^2+\cdots 
\end{equation}
For two dimensions $\sigma _1>0$ and $\sigma _2<0$. In this case one obtains
the following result:

{\bf \# There are scar states concentrated along the stable manifold of the
saddle point and, on the neighborhood of the stable manifold,}

\begin{equation}
\label{2.2}\left| \Psi _{\textnormal{scar}}\right| ^2\propto \cos \left( \frac{W_n%
}{2\hbar }x_2^2\right) \left| \psi _n(x_1)\right| ^2 
\end{equation}
$\psi _n(x_1)$ {\bf being close to an harmonic oscillator wave-function and }%
$W_n$ {\bf a function of the monodromy matrix in the transverse direction. }%
(Explicit expressions for $W_n$ under different approximations are given
below)

The result is obtained following Bogomolny's semiclassical construction of
wave functions\cite{Bogomolny}. From the series expansion of the energy
Green's function in terms of eigenfunctions it follows that the averaged
squared wave function is proportional to the imaginary part of the Green's
function 
\begin{equation}
\label{2.3}\left\langle \left| \Psi _{E_0}(x)\right| ^2\right\rangle \propto
\left\langle \textnormal{Im}G(x,x,E_0)\right\rangle 
\end{equation}
the average being taken over a small energy interval $\Delta E$ around $E_0$%
, which corresponds, in the semiclassical approximation, to restrict the
contributions to orbits with times of motion of order $\leq \frac \hbar
{\Delta E}$ . Likewise, if an average is taken over small intervals of the
variable $x$, the dominant contributions come from classical trajectories
for which the change of momentum on the closed orbit is small. I will
discuss later the role of these two averages.

The next step is to use the semiclassical approximation for the Green's
function 
\begin{equation}
\label{2.4}G(x_0,x,E)=\overline{G}(x_0,x,E)+\left( \frac 1\hbar \right)
^{(d+1)/2}G_{\textnormal{osc}}(x_0,x,E)
\end{equation}
\begin{equation}
\label{2.5}G_{\textnormal{osc}}(x_0,x,E)=i^{-1}\left( \frac 1{2\pi i}\right) ^{%
\frac{d-1}2}\sum_\beta \sqrt{\left| \det D_\beta \right| }\exp \left\{ \frac
i\hbar S_\beta \left( x_0,x,E\right) -i\frac \pi 2\nu _\beta \right\} 
\end{equation}
For the contributions in the neighborhood of a classical periodic orbit it
is convenient to choose one of the coordinates along the orbit $(x_1)$ and
the others $(x_i$ ; $i=2,\cdots ,n)$ along the transverse directions\cite
{Gutzwiller} \cite{Bogomolny}. On the neighborhood of an unstable periodic
orbit along the stable manifold of the saddle point the action is expanded
up to quadratic terms in the transversal coordinates and one has 
\begin{equation}
\label{2.6}G_{\textnormal{osc}}(x,x,E)=\frac 1{i\left( 2\pi i\right)
^{1/2}}\sum_\beta \frac{D^{1/2}(x_1)}{\left| \stackrel{\bullet }{x}_1\right| 
}\exp \left\{ \frac i\hbar \left( \overline{S}_\beta +\frac
12\sum_{i,j=2}^nW_{ij}(x_1)x_ix_j\right) -i\frac \pi 2\nu _\beta \right\} 
\end{equation}
with $D(x_1)$ and $W_{ij}(x_1)$ functions of the monodromy matrix in the
transverse coordinates 
\begin{equation}
\label{2.7}\left( 
\begin{array}{c}
x_{\bot }(\tau _1) \\ 
p_{\bot }(\tau _1)
\end{array}
\right) =\left( 
\begin{array}{cc}
m_{11}(x_1) & m_{12}(x_1) \\ 
m_{21}(x_1) & m_{22}(x_1)
\end{array}
\right) \left( 
\begin{array}{c}
x_{\bot }(0) \\ 
p_{\bot }(0)
\end{array}
\right) 
\end{equation}
with 
\begin{equation}
\label{2.8}
\begin{array}{c}
D(x_1)=\left| \det \left( m_{12}^{-1}\right) \right|  \\ 
W(x_1)=m_{12}^{-1}m_{11}+\left( m_{22}-1\right) m_{12}^{-1}-\left(
m_{12}^T\right) ^{-1}
\end{array}
\end{equation}
and $\tau _1$ the period of the periodic orbit along $x_1$. These
expressions hold for any number of transverse directions. I now specialize
to a two dimensional saddle point. Non-trivial orbits along the stable
manifold are harmonic motions of period $\tau _1=2\pi \sqrt{\frac{m_1}{%
2\sigma _1}}$ independent of the amplitude of the oscillation. Therefore
defining 
\begin{equation}
\label{2.9}D=\frac{\sqrt{2m_2\sigma _2}}{\sinh \left( 2\pi \sqrt{\frac{%
m_1\sigma _2}{m_2\sigma _1}}\right) }
\end{equation}
\begin{equation}
\label{2.10}W=\frac{2\cosh \left( 2\pi \sqrt{\frac{m_1\sigma _2}{m_2\sigma _1%
}}\right) -2}{\frac 1{\sqrt{2m_2\sigma _2}}\sinh \left( 2\pi \sqrt{\frac{%
m_1\sigma _2}{m_2\sigma _1}}\right) }
\end{equation}
$D$ and $W$ do not depend on $x_1$ and the dependence on the transverse
coordinate factors out for each term of the sum in Eq.(\ref{2.6}). However,
for each primitive trajectory, one also has to sum over multiple passings
obtaining the following sum 
\begin{equation}
\label{2.11}\sum_n\left( 2m_2\sigma _2\right) ^{\frac 14}\sinh ^{-\frac
12}\left( n\theta \right) \exp \left\{ \frac i\hbar \left( n\overline{S}+%
\frac{\cosh \left( n\theta \right) -1}{2m_2\sigma _2\sinh \left( n\theta
\right) }x_2^2\right) -i\frac \pi 2n\nu \right\} 
\end{equation}
where $\theta =2\pi \sqrt{\frac{m_1\sigma _2}{m_2\sigma _1}}$ . There are
two situations where simple closed form results may be obtained:

\# When $\exp \left( \theta \right) \gg 1$ , the $x_2$-dependence factors
out from the sum and the result is 
\begin{equation}
\label{2.12}\left\langle \left| \Psi _E(x)\right| ^2\right\rangle \propto
\left\langle \textnormal{Im}\left\{ \exp \left( \frac i\hbar \frac W2x_2^2\right)
G(x_1,E)\right\} \right\rangle 
\end{equation}
$G(x_1,E)$ being the harmonic oscillator Green's function. From 
\begin{equation}
\label{2.13}G(x_1,E)=\sum_n\left| \Psi _n(x)\right| ^2\left\{ P\left( \frac
1{E-E_n}\right) -i\pi \delta \left( E-E_n\right) \right\} 
\end{equation}
it follows that the principal part drops out under averages over small
energy intervals and the result (\ref{2.2}) follows with $W_n=W$ for $%
n=1,2,\cdots $ . The lowest lying state however corresponds to the orbit of
the unstable fixed point and the period of this orbit is no longer $\tau _1$%
. The value of $W_0$ in this case may be calculated by considering an orbit
from the coordinate $(0,x_2)$ to the fixed point $(0,0)$ and back along the
unstable manifold, which is equivalent to take the limit $\sigma
_1\rightarrow 0$ in Eq.(\ref{2.10}). Then 
\begin{equation}
\label{2.14}W_0=2\sqrt{2m_2\sigma _2} 
\end{equation}

\# If $\exp \left( \theta \right) $ is not large then the $x_2$-dependence
does not factor out in the sum (\ref{2.11}). Notice however that the sum in (%
\ref{2.11}) is not a sum over multiple passings of the same orbit, but a sum
over different orbits because, for $x_2\neq 0$ , the initial and final
momentum are different, and the difference grows with $n$. Therefore if, in
addition to the average over small energy intervals, one also averages over
small coordinate intervals then, the contribution of the primitive orbit
dominates the leading semiclassical approximation. The same factorization of
the $x_2$-dependence, as before, is obtained. However, the $\psi (x_1)$
function in this case is not exactly an harmonic oscillator wave functions,
but a function corresponding to a sum restricted to the primitive orbits.

\section{Applications}

The canonical form of the potential near a saddle point establishes a local
separation of variables which, of course, does not hold far away from the
saddle point. However, what the semiclassical estimate of the previous
section shows is that the local dynamics of the saddle point is sufficient
to insure the local existence of a factorized quantum state (\ref{2.2}). If
the separation of variables extends over a sufficiently large range then,
the transverse shape of the wave function may be approximated by the
solution of a one dimensional problem. Consider, for example a
one-dimensional Hamiltonian 
\begin{equation}
\label{3.1}H=-\frac 1{2m}\frac{d^2}{dx^2}+2g\cos (x) 
\end{equation}
with $2\pi -$periodic boundary conditions. The eigenstates are Mathieu
functions and there is a state of energy slightly above $2g$ with the
squared amplitude $\left| F\right| ^2$ as shown in Fig.1 (for $g=50$ , $%
\frac 1{2m}=1$ , $E=101.189$). In the figure the state is also compared with
the local approximation following from (\ref{2.2}) and (\ref{2.14}). The
width of the state depends on the factor $\sqrt{2mg}$ and the energy is
approximately $2g$ plus the kinetic localization energy. For an higher
dimensional problem one must also add the quantized energy of the harmonic
oscillation along the stable manifold. These considerations provide a simple
rule for the approximate energies at which saddle scars are expected to be
found.

Among the physical situations where saddle scars might appear, an
interesting example is probably the quantum collision of systems containing
both attractive and repulsive interactions (chemical ions, nuclei, etc)\cite
{Vilela}. When a system contains several positive and negatively charged
particles, there are classical configurations of close proximity of the
particles which are of low energy because the repulsion between the
like-charged particles is compensated by the attraction of unlike-charged
particles. These configurations however are highly unstable and the chance
to observe (or stabilize) them in classical mechanics is nil. They are zero
measure configurations in the energy surface. For smooth potentials these
unstable configurations would be saddle points of the potential, hence they
are expected to give rise to saddle scars. These states would correspond to
well defined energy levels and might be prepared by resonant excitation.
That is, quantum control through scars makes accessible some states that,
classically, are essentially unobservable.

Another interesting aspect of saddle scars is their generality, because
saddle points are the typical critical points of generic functions. There
are also other features of the classical phase space which the wave
functions imitate and that, in some limit or by change of coordinates, may
be related to the saddle scars. This concerns in particular the
regularization of singular potentials. An example is the collision states
found for a three-dimensional periodic Coulomb problem\cite{Vilela}. In this
case the quantum collision states correspond to wave functions concentrated
along a phase space feature which is not an actual orbit, but the separatrix
of two classes of unstable orbits. 

In Jacobi coordinates ($r=x_1-x_2$ , $\eta =x_3-\frac 12\left(
x_1+x_2\right) $) the potential between three unlike-charged particles is 
\begin{equation}
\label{3.2}V(r,\eta )=\frac 1{\left| r\right| }-\frac 1{\left| \frac r2-\eta
\right| }-\frac 1{\left| \frac r2+\eta \right| }
\end{equation}
The dynamics of binary collisions in the three-body problem may be
regularized, but the case of interest here is a triple collision which,
except for exceptional cases\cite{McGe}, is not regularizable. The
potential, however, may be regularized by addition of a small quantity to
the definition of the distances 
\begin{equation}
\label{3.3}\left| \rho \right| _\epsilon =\left( \rho _1^2+\rho _2^2+\rho
_3^2+\epsilon ^2\right) ^{\frac 12}
\end{equation}
In the neighborhood of the triple collision point $\overrightarrow{r}=%
\overrightarrow{\eta }=0$ , the regularized potential is 
\begin{equation}
\label{3.4}V_\epsilon =-\frac 1\epsilon -\frac 1{4\epsilon ^3}\left|
r\right| ^2+\frac 1{\epsilon ^3}\left| \eta \right| ^2+\cdots 
\end{equation}
and this is a saddle point with three stable and three unstable directions.
According to the discussion above one would expect the quantum collision
states to have scarred wave functions concentrated along the stable
manifolds and with a small dispersion in the transverse (unstable manifold)
directions. This is the situation that is indeed found in the numerical
computations\cite{Vilela} of the three-dimensional periodic Coulomb problem.
It shows that the effect seems to survive the $\epsilon \rightarrow 0$
limit. Triple collisions in a 3-dimensional 3-body problem lie on an
analytic 10-dimensional submanifold of the 12-dimensional dynamical
manifold. Hence it is a zero measure effect, essentially non-observable in
classical systems. It is interesting that, through the scar effect, they do
correspond to well-defined energy levels, accessible by resonant excitation.

\section{Figure caption}

Fig.1 - One dimensional density for a wave function concentrated around an
unstable point and the semiclassical approximation (+).

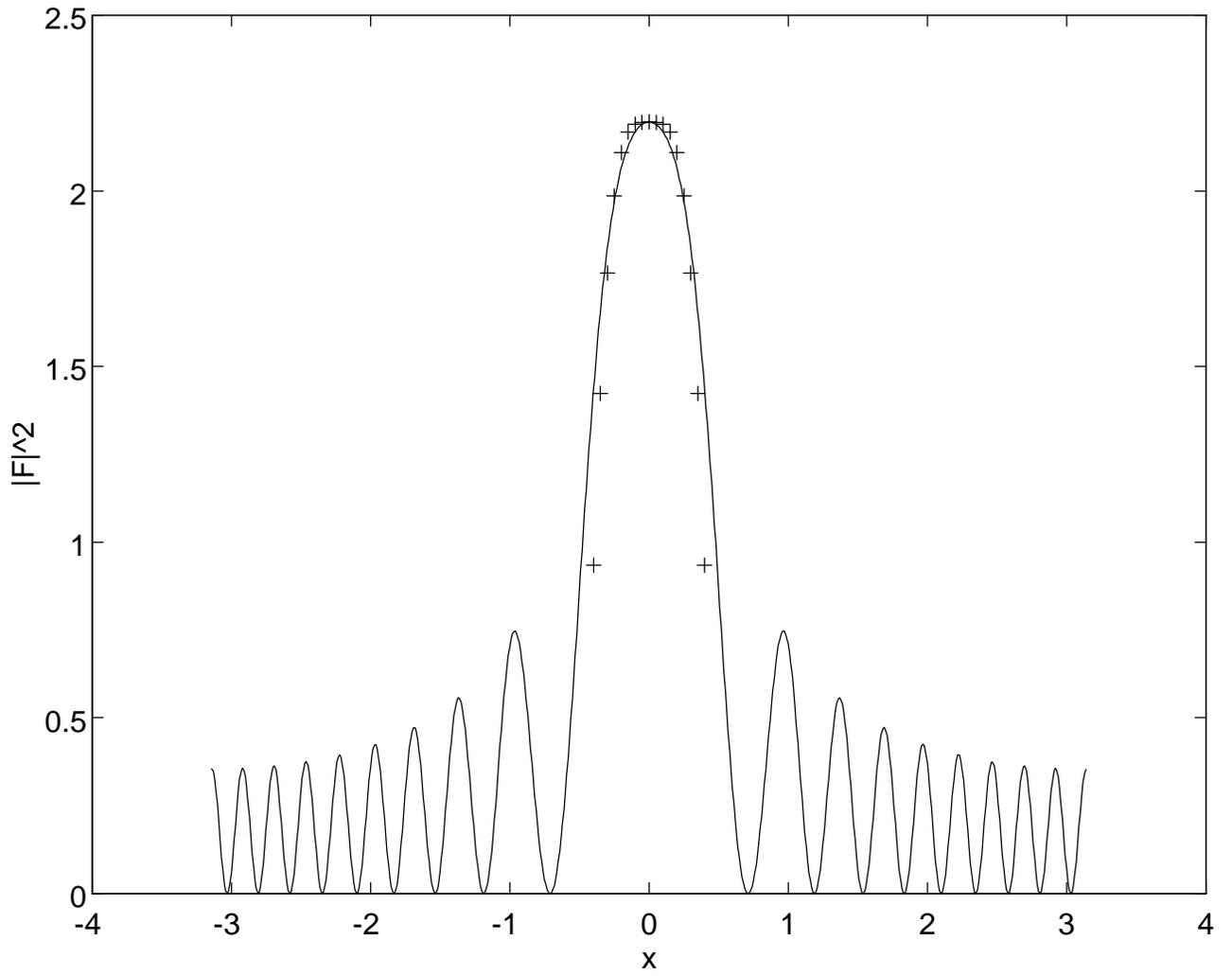
\end{document}